\documentclass[12pt,fleqn]{article}
\usepackage{amsfonts}

\begin{document}

\title{Space, Time and Coordinates in a Rotating World}

\author{Dennis Dieks\\ Institute for the History and Foundations of Science\\
Utrecht University, P.O.Box 80.010 \\ 3508 TA Utrecht, The
Netherlands\\Email: d.dieks@uu.nl}
\date{}
\maketitle

\begin{abstract}
The peculiarities of rotating frames of reference played an
important role in the genesis of general relativity. Considering
them, Einstein became convinced that coordinates have a different
status in the general theory of relativity than in the special
theory. This line of thinking was confused, however. To clarify
the situation we investigate the relation between coordinates and
the results of space-time measurements in rotating reference
frames. We argue that the difference between rotating systems (or
accelerating systems in general) and inertial systems does not lie
in a different status of the coordinates (which are conventional
in all cases), but rather in different global chronogeometric
properties of the various reference frames. In the course of our
discussion we comment on a number of related issues, such as the
question of whether a consideration of the behavior of rods and
clocks is indispensable for the foundation of kinematics, the
influence of acceleration on the behavior of measuring devices,
the conventionality of simultaneity, and the Ehrenfest paradox.
\end{abstract}

\newpage

\section{Introduction}\label{introduction}
In his Autobiographical Notes \cite{einstein}, Einstein relates
how important Machian empiricist ideas were for his discovery of a
theory that could reconcile the idea that all inertial frames are
equivalent with the principle that the velocity of light has a
fixed value that is independent of the velocity of the emitting
source. It was essential, he states, to realize what the meaning
of \emph{coordinates} in physics is: they are nothing but the
outcomes of length and time measurements by means of rods, clocks
and light signals. This idea led Einstein to his famous critique
of the classical notion of simultaneity, one of the cornerstones
of the special theory of relativity.

It soon turned out, however, that the special theory of relativity
was not able to accommodate gravitation, and the principle of
equivalence, in a natural way. Einstein fully recognized this
problem in 1908, but it took him another seven years before he
succeeded in constructing the general theory. As he explains in
his Autobiographical Notes, the main reason for the slowness of
his progress in this period was the difficulty of
\emph{abandoning} again, in the context of the general theory, the
idea that coordinates should possess immediate metrical meaning.

From a systematical (as opposed to a historical or psychological)
point of view this emphasis on the different meaning of
coordinates, in the context of the two theories, is very odd. For
the practice of physics before, during and after Einstein's days,
even if governed by the severest empiricist norms, does not at all
indicate that coordinates should possess a metrical significance,
relating to the indications of rods and clocks. Think, for
example, of the way coordinates are used in observational
astronomy: the essential thing is that the coordinates are
assigned to celestial objects in an objective and reproducible
way; how the coordinates relate to distances is a matter to be
found out subsequently. Coordinates are even routinely attributed
to regions of the universe in which rods and clocks could not
possibly exist. This is obviously unobjectionable from an
empiricist point of view, as long as the method by which the
coordinates are assigned is operationally specified. So, even
within the framework of special relativity general coordinate
systems that do not reflect the indications of rods and clocks are
entirely permissible.

What finally led Einstein to abandon his special relativistic
analysis of the meaning of coordinates, he tells us, was the lack
of metrical significance of coordinates in accelerating frames of
reference; the consideration of coordinates on a rotating disc
played an important role in reaching this conclusion
\cite{stachel}. But, as we will see, there is confusion here: the
metrical significance of coordinates in accelerating frames can be
determined completely through application of the principles of
\textit{special} relativity, so there can be no need to revise the
meaning of the notion of coordinates, or to invoke a new
epistemological analysis.

As it turns out, the difference between inertial and non-inertial
frames of reference, and between special and general relativity,
is not in the epistemological status of the coordinates. Rather,
the difference is that chronogeometric characteristics become
globally different. This is a physical rather than a philosophical
difference, and has nothing to do with the meaning or
permissibility of coordinate systems.

The rotating frame of reference nicely illustrates these points.
There is no problem in defining operationally meaningful
coordinates in a rotating (and therefore accelerating) frame.
Furthermore, relating these coordinates to distances and time
intervals, and the behavior of moving objects, can be done by the
means provided by special relativity. However, the spatial
geometry becomes non-Euclidean, and local Einstein synchrony does
not lead to a global notion of time. These latter features
constitute the essential differences from the situation in an
inertial frame.

In the course of our discussion we will have occasion to comment
on a number of related issues, such as the status of rods and
clocks, the behavior of accelerating measuring devices, the
conventionality of simultaneity, and the Ehrenfest paradox.

\section{The rotating frame of reference}\label{frame}

Let us start from Minkowski space-time, coordinatized by inertial
coordinates $r$, $\varphi$, $z$ and $t$: $r$ and $\varphi$ are
polar coordinates in a plane, $z$ is a Cartesian coordinate
orthogonal to this plane, and $t$ is the standard time coordinate.
It so happens that $r$, $z$, and $t$ can be thought of as
representing the indications of rods and clocks, but that is not
important for their role as coordinates, which is just to pinpoint
events unequivocally. The choice of coordinates is conventional
and pragmatic. In this case we choose polar coordinates because we
are going to describe a system that possesses axial symmetry:
polar coordinates simplify the description.

Once we have laid down coordinates, the metrical aspects should be
introduced via further stipulations. This is ordinarily done
through the introduction of the `line element' $ds^2 = c^2 dt^2
-dr^2 - r^2 d\varphi^2-dz^2$, plus a specification of what this
mathematical expression represents physically. The traditional
approach is to invoke standard rods and clocks: $ds/c$ is the time
measured by a standard clock whose $r$, $\varphi$ and $z$
coordinates are constant. Furthermore, $\sqrt{-ds^2}$ is the
length of a rod with a stationary position in the coordinates and
with constant coordinates and differences $dr$, $d\varphi$, $dz$
between its endpoints, taken at one instant according to standard
simultaneity ($dt=0$). However, it would be a mistake to think
that rods and clocks are indispensable to relate the coordinates
to metrical concepts. In section \ref{without} below we will
discuss an approach that does not make use of rods and material
clocks.

We now introduce alternative coordinates for the events in this
Minkowski world: $t^\prime = t$, $r^\prime = r$, $\varphi^\prime =
\varphi - \omega t$ and $z^\prime = z$, with $\omega$ a constant.
Since rest in the new coordinates obviously means uniform rotation
with respect to the old frame, we call the frame of reference
defined by these new coordinates the \textit{rotating frame of
reference}.

It is clear that if operational methods are at hand to fix the old
coordinates, the same methods can be used to assign values to the
new coordinates (we assume $\omega$ to be known). So from an
empiricist or operational point of view the new coordinates are
impeccable. However, from the special theory of relativity we know
that material bodies at rest in the new coordinates may not exist
($\omega r$ may be greater than $c$, the velocity of light). It is
true, therefore, that the new coordinates will not always have a
direct interpretation in terms of co-moving bodies---but this is
something to be distinguished sharply from the more general
question of whether they have adequate empirical significance at
all.

Substitution of the rotating coordinates into the expression for
the line element yields $ds^2 = (c^2 - {r^\prime}^2 \omega^2)
{dt^\prime}^2 -{dr^\prime}^2 - {r^\prime}^2
{d\varphi^\prime}^2-{dz^\prime}^2 - 2\omega {r^\prime}^2
d\varphi^\prime dt^\prime $. As we already mentioned, it is a
basic principle of the special theory of relativity that the line
element supplies all information about the physics of the
situation, as described in the given coordinates. It was also
mentioned above that the traditional link between $ds$ and
physical concepts makes use of clocks and measuring rods. However,
there is another and more fundamental physical interpretation
available that only makes use of the basic laws of motion: as long
as no disturbing forces act, point particles follow time-like
geodesics and light follows null-geodesics in the metric defined
by $ds^2$. The relation between these dynamical aspects (how
particles and light move) and the metrical aspects (rods and
clocks) will be the subject of comments in section \ref{without}.

\section{Rods and clocks}\label{rods}

Let us for the moment stay with the physical interpretation of
$ds$ in terms of measurements performed with rods and clocks.
Concerning time, the coordinating principle is that $ds/c$
represents proper time, measured by a clock whose world line
connects the events between which $ds$ is calculated. This
principle entails that a clock at rest in the rotating frame will
indicate the proper time
\begin{equation}\label{time}
 ds/c= \sqrt{(1 - {r^\prime}^2
\omega^2 /c^2 )} dt^\prime.
\end{equation}
Because $t^\prime = t$ and $t$ has the physical meaning of the
time indicated by a clock at rest in the old frame, this implies
that clocks at rest in the rotating frame are slow compared to
clocks in the original (``laboratory'') frame.

With regard to spatial distances, the interpretative principle is
that $\sqrt{-ds^2}$ gives the length of an infinitesimal rod whose
endpoints are simultaneous according to standard simultaneity in
the rod's rest frame (\cite{reichenbach}, p.187). (A rod is a
three-dimensional object, so we need a stipulation about the
instants at which its endpoints should be considered in order to
get a four-dimensional interval for which $ds$ can be calculated.)
When we apply this rule to rods that are at rest in the rotating
frame of reference, we encounter the complication that $dt^\prime
=0$ does not automatically correspond to standard simultaneity in
the rotating frame. The definition of standard synchrony of two
(infinitesimally near) clocks A and B is that a light signal sent
from A to B and immediately reflected to A, reaches B when B
indicates a time that is halfway between the instants of emission
and reception, respectively, as measured by A. Suppose that A and
B, both at rest in the rotating frame, have positions with
coordinate differences $dr$, $d\varphi$ and $dz$---from now on we
drop the primes of the rotating coordinates. A light signal
between A and B follows a null-geodesic:
\begin{equation}\label{null}ds^2=(c^2 - {r}^2 \omega^2) {dt}^2 -{dr}^2 -
{r}^2 {d\varphi}^2-{dz}^2 - 2\omega {r}^2 d\varphi dt =
0.\end{equation} This equation gives the following solutions for
$dt$ when it is applied to the signals from A to B and back,
respectively:
\begin{equation}\label{roundtrip}
dt_{1,2}= \frac{\pm \omega r^2 d\varphi + \sqrt{(c^2-\omega^2 r^2)
(dz^2 + dr^2) + c^2 r^2 d\varphi^2}}{c^2 -\omega^2
r^2}.\end{equation} If $t_0$ is the time coordinate of the
emission event at A, the event at A with time coordinate $t_0 +
1/2(dt_1 + dt_2)$ is standard-simultaneous with the event at B
with time coordinate $t_0 + dt_1$. It follows that standard
synchrony between infinitesimally close events corresponds to the
following difference in $t$-coordinate:
\begin{equation}\label{synchrony}
 dt= (t_0 + dt_1) - (t_0 + 1/2dt_1 + 1/2dt_2)=(\omega r^2 d\varphi)/(c^2-\omega^2 r^2).\end{equation} As was to
be expected, it is only for events that differ in their
$\varphi$-coordinates that $dt=0$ is not equivalent to standard
simultaneity; indeed, the instantaneous velocity of the rotating
frame is tangentially directed, and the relativistic dilation and
contraction effects only take place in the direction of the
velocity.

The spatial distance between two infinitesimally near points, as
measured by a rod resting in the rotating frame, is found by
substituting the just-derived value of $dt$, (\ref{synchrony}), in
the expression for $ds$. The result is the following expression
for the 3-dimensional spatial line element:
\begin{equation}\label{space}
dl^2 = dr^2 + \frac{r^2d\varphi^2}{1-\omega^2 r^2/c^2} + dz^2.
\end{equation}

We could have found (\ref{time}) and (\ref{space}) in a simpler
way by making use of the standard expressions for the time
dilation and Lorentz contraction undergone by clocks and rods,
respectively, that possess the instantaneous velocity $\omega r$.
However, the use of the line element as the central theoretical
quantity provides us with a unifying framework that makes it
easier to discuss the relation between metrical and dynamical
concepts.

\section{Space and time without rods and clocks}\label{without}

In his Autobiographical Notes, Einstein already points out that
from a fundamental point of view it is unsatisfactory to interpret
$ds$ via measuring procedures with complicated macroscopic
instruments. Indeed, this could create the false impression that
rods and clocks are basic entities without which the theory would
have no physical content. However, it is clear that rods and
clocks themselves consist of more fundamental entities, like atoms
and molecules. In principle it would therefore be better to base
the interpretation of the theory directly on what it says about
the fundamental constituents of matter. It is only because no
complete theory of matter was available, Einstein explains, that
it was expedient to introduce the theory through measurements by
rods and clocks. In principle they should be eliminated at a later
stage.

This desideratum, to do without rods and clocks, becomes even more
urgent when accelerated frames of reference are considered, as in
the case of our rotating world. Obviously the motions of clocks
and rods that are stationary in the rotating frame are not
inertial. Centrifugal and Coriolis forces will therefore arise,
which will distort the rotating instruments. It is not a priori
clear that such deformed instruments will keep on functioning as
indicators of $ds$. Indeed, one could easily think of rods or
clocks that would be torn apart by centrifugal forces and would
therefore certainly not indicate any length or time intervals.

Fortunately, it \textit{is} possible to found the space-time
description of our rotating world on a more fundamental level than
that of macroscopic measuring devices. In fact, in general
space-times one can use the basic principles that time-like
geodesics are physically realized by inertially moving
point-particles and that null-geodesics represent light rays, to
define space-time distances between neighboring events
(\cite{wheeler}, section 16.4). In our case, Minkowski space-time,
we can start by constructing a set of elementary `light clocks' by
letting light signals bounce back and forth between neighboring
parallel particle geodesics. If we confine our attention to the
plane $z=0$, we can take the geodesics defined in the laboratory
frame (the inertial system we started with) by constant $r$,
$\varphi$ and $r+dr$, $\varphi$, respectively. The thus
constructed clock has a constant period (the $dt$ between two
`ticks') of $2dr/c$. In other words, we have here an elementary
process that provides a physical realization of $t$; and we have
come to this conclusion on the basis of the dynamical postulates
alone (the only ingredient is that light follows null-geodesics).
Length can be determined in a similar way: let a light signal
depart from A, with fixed $r$ and $\varphi$ and go to a
neighboring position B with $r+dr$ and $\varphi + d\varphi$ from
which it returns immediately to A. Let the round trip time
measured at A be $dt$. We can now define the spatial distance $dl$
between A and B as $c dt/2$. From the postulate that light follows
null-geodesics it follows that $dl^2 = dr^2 + r^2 d\varphi^2$. In
this way the laboratory coordinates obtain metrical significance,
without reliance on macroscopic clocks and rigid rods. When such
(complicated) systems are introduced at a later stage, we can
study their workings on the basis of the fundamental laws of
physics governing their constituents and see, on that basis,
whether they are indeed suitable to measure the just-defined
intervals.

We now turn our attention to measurements performed within the
rotating system, i.e.\ with instruments resting in the rotating
coordinates. From Eq.\ (\ref{roundtrip}) we see that the round
trip time $dt$ needed by a light signal between two neighboring
points that are stationary in the rotating frame of reference is
given by
\[ dt= dt_1 +dt_2= 2 \frac{\sqrt{(c^2-\omega^2 r^2)dr^2 + c^2 r^2
d\varphi^2}}{c^2 -\omega^2 r^2}.\] If the laboratory coordinate
$t$ is used as the measure of time, and if the definition
$dl=cdt/2$ is used to fix spatial distances, we arrive at the
metric \[dl^2= \frac{(1-\omega^2 r^2/c^2)dr^2 +
r^2d\varphi^2}{(1-\omega^2 r^2/c^2)^2}.\] However, it is more
natural to link the measure of time intervals in the rotating
system to the indications furnished by light clocks that are
co-moving, i.e.\ stationary in the rotating coordinates instead of
stationary in the laboratory frame. So let a light ray bounce back
and forth between two points that only differ in their
$r$-coordinate, by the amount $dr$, in the rotating frame. It
follows from the expression (\ref{null}) that the period of the
thus defined clock is $2dr/\sqrt{c^2-\omega^2r^2}$, whereas the
period of the similar and instantaneously coinciding clock in the
laboratory frame is $2dr/c$. The period of the rotating light
clock is therefore longer, by a factor
$1/\sqrt{1-\omega^2r^2/c^2}$, than the period of the laboratory
clock. When we now define distances as $cd\tau/2$, with $\tau$
measured in the new `co-moving' time units, we have to multiply
the distances we found a moment ago by $\sqrt{1-\omega^2
r^2/c^2}$. The final result is \[dl^2= dr^2 +
r^2d\varphi^2/(1-\omega^2 r^2/c^2).\] This is the same result as
we found in Eq.\ (\ref{space}).

\section{Accelerating measuring devices}\label{accelerating}

The above sketch shows how we can achieve a physical
implementation of the two systems of coordinates, and give them
metrical meaning, by the sole use of point-particles and light.
The thus defined space-time distances can be used to calibrate
macroscopic measuring rods and clocks. Indeed, it is clear that in
general such instruments will be deformed by the rotational
motion, and that this will introduce inaccuracies in their
readings.

The general effect of accelerations can be illustrated by the
consideration of a light-clock of the kind mentioned above: a
light signal bouncing back and forth between two particle
world-lines. Light travelling to and fro between two mirrors
resting in an inertial system, with mutual distance $L$, defines a
clock with half period $T=L/c$. When the two mirrors move
uniformly with the same velocity $\overrightarrow{v}$, in a
direction parallel to their planes, a simple application of the
Pythagorean theorem shows that the half period of the moving clock
becomes $L/(c\sqrt{1-v^2/c^2})=T/\sqrt{1-v^2/c^2}$. This
demonstrates the presence of time dilation in the case of a moving
light-clock (by means of the relativity principle this result can
be extended to other time-keeping devices). Consider now what
happens if the velocity is not uniform but the system starts
accelerating when the light leaves the first mirror, with a small
acceleration $\overrightarrow{a}$ in the direction of
$\overrightarrow{v}$. As judged from the inertial frame, the light
now needs a time $T^\prime$ to reach the second mirror; during
this time the accelerating mirror system has covered a distance $s
\approx vT^\prime + 1/2a{T^\prime}^2$. Application of Pythagoras
now yields $c^2 {T^\prime}^2 = L^2 + s^2$. It follows that
\begin{equation}\label{period}
c^2 {T^\prime}^2 = L^2 + v^2 {T^\prime}^2 +av {T^\prime}^3 + 1/4
a^2 {T^\prime}^4.
\end{equation}
The half period ${T^\prime}$ that follows from this equation
obviously depends on $a$. However, it is also obvious that the
extent of the change in the period caused by $a$ depends on the
magnitude of ${T^\prime}$ itself. If we make $T^\prime$ in Eq.\
(\ref{period}) very small, by reducing $L$, we find in the
limiting situation $T^\prime = T/\sqrt{1-v^2/c^2}$, just as in the
case of the uniformly moving clock. In other words, the
acceleration has an effect, but the magnitude of this effect
depends on the peculiarities of the specific clock we are
considering (in this case on $L$). This acceleration-dependent
effect can be made as small as we wish, by using suitably
constructed clocks (in the example: by reducing $L$). What remains
in all cases is the universal effect caused by the velocity.

This shows in what sense velocities have a universal effect on
length and time determinations, but accelerations not. There is no
independent postulate involved here; everything can be derived
from the dynamical principles of special relativity theory, by
considering the inner workings of the measuring devices. It turns
out that acceleration-dependent effects are there, but can be
varied, and corrected for, by varying the characteristics of the
devices. This is the real content of the textbook statement that
acceleration has no metrical effects. It should be stressed again
that this does not constitute a new hypothesis that has to be
\textit{added} to the dynamical principles of the theory of
relativity. Quite to the contrary, the effects of accelerations on
any given clock or measuring rod can be computed from the
dynamical principles applied to these devices.

Of course, that the magnitudes of distortions will depend on the
specific constitutions of the rods or clocks in question is only
to be expected. Robust rods and clocks will be less affected
accelerations than fragile ones. One way of correcting for the
deformations is to gauge the accelerating instruments against the
light measurements results described in section (\ref{without}).
The expressions (\ref{time}) and (\ref{space}) should be
understood as applying to the results of space-time measurements
performed with thus corrected measuring devices.

\section{Space and time in the rotating frame}\label{rotspacetime}

The spatial geometry defined by the line element (\ref{space}) is
non-Euclidean, with a negative $r$-dependent curvature (see
\cite{tonnelat}, pp.\ 330-337). One of the notorious
characteristics of this geometry is that the circumference of a
circle with radius $r$ (in the plane $z =0$) is $2\pi r
/(1-\omega^2 r^2/c^2)$, which is greater than $2\pi r$. The
recognition that the geometry in accelerated frames of reference
will in general be non-Euclidean, which through the equivalence
principle suggests that the presence of gravitation will also
cause deviations from Euclidean geometry, played an important role
in Einstein's route to General Relativity. We will restrict
ourselves to the special theory, however.

The properties of time in the rotating frame are perhaps even more
interesting than the spatial characteristics. Expression
(\ref{synchrony}) demonstrates that standard simultaneity between
neighboring events in the rotating frame corresponds to a non-zero
difference $dt$. It follows that if we go along a circle with
radius $r$, in the positive $\phi$-direction, while establishing
standard simultaneity along the way, we create a `time gap'
$\triangle t = {2\pi \omega r^2}/(c^2-\omega^2 r^2)$ upon
completion of the circle. Doing the same thing in the opposite
direction results in a time gap of the same absolute value but
with opposite sign. So the total time difference generated by
synchronizing over a complete circle in one direction, and
comparing the result with doing the same thing in the other
direction is $\triangle t = {4\pi \omega r^2}/(c^2-\omega^2 r^2).$

Now suppose that two light signals are emitted from a source fixed
in the rotating frame and start travelling, in opposite
directions, along the same circle of constant $r$. We follow the
two signals while locally using standard synchrony; this has the
advantage that locally the standard constant velocity $c$ can be
attributed to the signals. We therefore conclude that the two
signals use the same amount of time in order to complete their
circles and return to their source, as calculated by integrating
the elapsed time intervals measured in the successive local
comoving inertial frames (the signals cover the same distances,
with the same velocity $c$, as judged from these frames). However,
because of the just-mentioned time gaps the two signals do not
complete their circles simultaneously, in one event. There is a
time difference $\triangle t = {4\pi \omega r^2}/(c^2-\omega^2
r^2)$ between their arrival times, as measured in the coordinate
$t$. This is the celebrated Sagnac effect (see \cite{nienhuis},
p.\ 652 for a related derivation).

The Sagnac effect directly reflects the space-time geometry of the
rotating frame; it does not depend on the specific nature of the
signals that propagate in the two directions. Indeed, as long as
the two signals have the same velocities in the locally defined
inertial frames with standard synchrony, the difference in arrival
times is given by the above time gap. So the same Sagnac time
difference is there not only for light, but for any two identical
signals running into two directions. The Sagnac experiment
directly probes the space-time relations in the rotating frame.

Because of the difference in arrival times of the two light
signals, the velocity of light obviously cannot be everywhere the
same in the rotating coordinates. This is a consequence of the
fact that in the rotating frame events with equal time coordinate
$t$ are not standard simultaneous. So $t$ may appear as an
unnatural time coordinate for the rotating frame: it would be
desirable to have a time coordinate that \textit{would} reflect
standard simultaneity everywhere. The question can therefore be
asked whether we could define a coordinate $\tilde{t}$ in such a
way that $d\tilde{t}=0$ would imply standard synchrony in the
local inertial frame. Suppose that
$\tilde{t}=\tilde{t}(t,r,\varphi)$, then we should have that
$d\tilde{t}=0$ if Eq.\ (\ref{synchrony}) holds. This implies that
${\omega^2 r^2/(c^2 - \omega^2 r^2)}
\partial\tilde{t}/{\partial t}  +
\partial\tilde{t}/\partial\varphi = 0$ and
$\partial\tilde{t}/\partial r = 0.$ In view of the axial symmetry
in our frame we may assume that $\partial\tilde{t}/\partial\varphi
= 0. $ The only solution of our partial differential equations is
therefore that $\tilde{t}$ is independent of $r$, $\varphi$ and
$t$, which clearly is unacceptable. Therefore, it turns out to be
a basic characteristic of the rotating frame that the locally
defined Lorentz frames do not mesh: they cannot be combined into
one frame with a globally defined standard simultaneity. Evidently
it \textit{is} possible to define global time coordinates, like
$t$; but the description of physical processes in terms of these
coordinates must necessarily differ from the standard description
in inertial systems. The non-constancy of the velocity of light in
the rotating system furnishes an example. It should be noted that
this peculiarity of the description of physical processes in the
rotating system is not a consequence of the presence of
centrifugal and Coriolis forces: indeed, in our space-time
determinations we have compensated for the effects of such forces.
It is the space-time geometry itself that is at issue.

\section{Simultaneity, slow clock transport and conventionality}

As we saw in the previous section, the Sagnac effect is
independent of the nature of the signals that propagate into the
two directions on the rotating disc. So, if we transport two
clocks along a circle with radius $r$ around the center of the
disk, one clockwise and one counter-clockwise, while keeping their
velocities the same in the locally co-moving inertial frames,
there will be a difference $\triangle t = {4\pi \omega
r^2}/(c^2-\omega^2 r^2)$ between their return times (measured in
the laboratory time $t$). It is well known that the indications of
the clocks will conform to standard simultaneity in the limiting
situation of vanishing velocities. That is, if the clocks are
transported very slowly with respect to the rotating disc, they
will remain synchronized according to standard simultaneity in the
local inertial frames. It follows that slow clock transport cannot
be used to define an unambiguous global time coordinate on the
rotating disc: in the just-mentioned case the result will depend
on whether a clockwise or counter-clockwise path is chosen. In
general, the result of synchronization by slow clock transport
will be path dependent.

With regard to time in inertial frames, there has been a
long-standing and notorious debate about whether standard
simultaneity ($\varepsilon=1/2$ according to Reichenbach's
formulation) is conventional or not. One of the arguments often
put forward against the conventionality thesis is that the natural
procedure of slow clock transport leads to $\varepsilon=1/2$, thus
showing its privileged status. In the case of the rotating world,
this argument can only be applied locally. Neither the Einstein
light signal procedure, nor the slow transport of clocks can be
used to establish a global notion of simultaneity on the rotating
disc.

More generally, it cannot be denied that in inertial frames
standard simultaneity has a special status: it allows a simple
formulation of the laws, conforms to slow clock transport and
other physically plausible synchronization procedures, and agrees
with Minkowski-orthogonality with respect to world lines
representing the state of rest \cite{malament}. So time
coordinates $t$ that correspond to this notion of simultaneity (in
the sense that $dt = 0$ expresses simultaneity) may be said to be
privileged. In non-inertial frames this still is so, though now
the argument applies only locally. The rotating system illustrates
the situation very well: in each point on the disc standard
simultaneity can be defined just as in an inertial system, but
this does not result in a global time coordinate. This supports
the general conclusion of this paper, namely that the difference
between the status of coordinates in inertial and non-inertial
frames of reference, or special and general relativity, is not so
much a matter of epistemology---or philosophical analysis of the
meaning of coordinates---but rather a matter of physical facts. In
global inertial systems privileged coordinates can be chosen that
have a global metrical interpretation. In reference frames that
are not globally inertial such privileged coordinates do not exist
in general. This is not a matter of a different philosophical
status of coordinates, but rather a reflection of different global
space-time symmetry properties---a factual physical difference
rather than a philosophical distinction.

The purpose of \emph{coordinates} is to label events
unambiguously, which can be done in infinitely many different
ways. The choice between these different possibilities is a matter
of pragmatics; though there may be very good reasons to prefer one
choice over another. Thus, in inertial frames of reference time
coordinates that reflect standard simultaneity lead for many
purposes to an especially simple description. In this case there
exists a physically significant global temporal relation between
events, and coordinates that are adapted to this relation inherit
its special status. But in the general case no physically
significant simultaneity relation exists. Global "simultaneity"
can then
 only refer to some global time coordinate, which is
chosen conventionally. This is true in non-inertial frames of
reference, like the rotating disc, and in generally relativistic
space-times in which there are no global temporal symmetries.
These non-inertial frames of reference, and general relativistic
space-times, seem an arena where the thesis that (global)
simultaneity is conventional can be defended without controversy.

\section{The rotating Ehrenfest cylinder}\label{ehrenfest}

Not only in its temporal aspects, but also in its spatial physical
properties the rotating frame differs globally from an inertial
frame. Until now we spoke about a rotating frame of reference as
defined by a set of rotating \emph{coordinates}, without
discussing a possible material realization of this frame. It is
clear from the outset that the special theory of relativity sets
limits to such a realization: objects at rest in the rotating
frame should not move faster than light as judged from the
inertial laboratory frame. This implies that $\omega r < c$ should
hold for such an object. In other words, there is an upper bound
to the value of $r$ that can be realized materially.

However, even if this condition is satisfied there remain
interesting questions, as made clear by Ehrenfest in his famous
note on the subject \cite{paradox}. Suppose that a solid cylinder
of radius $R$ is gradually put into rotation about its axis;
finally it reaches a state of uniform rotation with angular
velocity $\omega$. It would seem that in the final state the
cylinder has to satisfy contradictory requirements: on the one
hand the Lorentz contraction should make the circumference
shorter, on the other hand the radial elements should not contract
because their motion is normal to their lengths. From symmetry it
is clear that the form of a cross section of the moving cylinder
remains a circle, as judged from the laboratory frame; but this
would apparently mean that the circumference of the circle has
become smaller while the radius has stayed the same. This is
inconsistent (remember that Euclidean geometry holds in the
laboratory frame).

The solution of this paradox is that the various parts of the
cylinder, being fastened to each other, cannot move freely and
therefore cannot Lorentz contract as freely moving infinitesimal
measuring rods would do. What will happen to the cylinder during
its acceleration depends on the elastic properties of the
material: tensions will develop because the tangential elements
want to shrink, whereas the radial elements do not. A possible
scenario is that the tangential elements will be stretched as
compared to their natural (i.e.\ Lorentz contracted) lengths.
Another possibility, if the material is sufficiently strong, is
that the radius will contract, allowing the circumference to
contract too. However, if $\omega$ becomes big enough one would
have to expect that the tensions and strains grow to such an
extent that they cause the cylinder to explode. This makes it
clear that the Lorentz contraction can be responsible for clearly
dynamical effects---the contractions are not just a matter of
``perspective'' (see \cite{dieks1} and \cite{dieks2}). (Of course,
this whole discussion is rather academical because centrifugal
forces will tear the cylinder apart before the relativistic
effects become noticeable.)

As long as the cylinder survives, and keeps its cylindrical shape
(as judged from the laboratory frame), not all its elements will
be free from deformations, tensions or strains. However, the
length determinations by measuring rods at rest in the rotating
frame, as discussed in section \ref{rods}, were supposed to be
carried out with freely movable rods that are not hampered in
their Lorentz contractions. So measuring rods laid out along the
circumference of the circle will have undergone a Lorentz
contraction, whereas rods laid out along a radius will have
retained their rest length (as judged from the laboratory system).
The measurement would reveal that the circumference is longer than
$2\pi$ times the radius, in conformity with equation
(\ref{space}).

The spatial geometry of the disc is therefore non-Euclidean. That
means that distance relations must be represented by a metrical
tensor that cannot be put into the Euclidean diagonal form
everywhere. It remains possible, of course, to choose coordinates
locally in such a way that the Euclidean form results at the point
in question. The difference from the inertial system concerns
global aspects, not local ones. The impossibility to define a
global coordinate system in which the metrical tensor reduces to
its Euclidean standard form implies that there cannot be
coordinates whose differences correspond to distances everywhere.
The situation is analogous to the one we discussed in the context
of time coordinates: nothing changes in the status and meaning of
coordinates when we go from inertial to non-inertial systems. The
things that do change are the global characteristics of the
physical geometry, which are coordinate-independent.

\section*{Conclusion}

The transition from inertial to non-inertial frames of reference,
and the transition from special to general relativity, does not
imply a change in the status and meaning of coordinate systems. It
is therefore a misunderstanding to think that general relativity
allows a wider class of coordinate systems than classical physics
or special relativity. In classical physics and in relativity
theory, both in inertial systems and non-inertial systems,
coordinates just serve to label events. The choice for a
particular coordinate system from the infinity of possible ones is
dictated by pragmatic considerations.

What \emph{does} change in the transition from inertial to
non-inertial systems, and from special to general relativity, are
the global aspects of the physical spatial and temporal relations.
Pragmatic arguments for choosing one coordinate system over
another may therefore lead to different choices in the different
situations: if geometrical relations have become different,
coordinate systems with different characteristics, adapted to the
new geometry, may lead to a simpler description. But this does not
change the conventional nature of the coordinates.

\end{document}